\documentclass[letterpaper, 10 pt, conference]{ieeeconf}  

\IEEEoverridecommandlockouts                              

\usepackage{graphicx}
\usepackage{xcolor}
\usepackage{array}
\usepackage[bookmarks=false]{hyperref}
\newcolumntype{x}[1]{>{\centering\let\newline\\\arraybackslash\hspace{1pt}}p{#1}}

\newcommand{\specialcell}[2][c]{\begin{tabular}[#1]{@{}c@{}}#2\end{tabular}}
\newcolumntype{M}[1]{>{\centering\arraybackslash}m{#1}}
\newcommand\norm[1]{\left\Vert#1\right\Vert}
\PassOptionsToPackage{hyphens}{url}\usepackage{hyperref}
\PassOptionsToPackage{bookmarks=false}{hyperref}
\usepackage{etoolbox}
\makeatletter
\patchcmd{\@makecaption}
  {\scshape}
  {}
  {}
  {}
\makeatother

\usepackage{tikz}
\usepackage{textcomp}
\usepackage{hyperref}
\usepackage{lipsum}

\newcommand\copyrighttext{%
  \footnotesize \textcopyright 2020 IEEE.  Personal use of this material is permitted.  Permission from IEEE must be obtained for all other uses, in any current or future media, including reprinting/republishing this material for advertising or promotional purposes, creating new collective works, for resale or redistribution to servers or lists, or reuse of any copyrighted component of this work in other works.}
\newcommand\copyrightnotice{%
\begin{tikzpicture}[remember picture,overlay]
\node[anchor=south,yshift=10pt] at (current page.south) {\fbox{\parbox{\dimexpr\textwidth-\fboxsep-\fboxrule\relax}{\copyrighttext}}};
\end{tikzpicture}%
}

\title{\LARGE \bf
EEG Channel Interpolation Using Deep Encoder-decoder Networks
}

\author{Sari Saba-Sadiya$^{1}$, Tuka Alhanai$^{2}$, Taosheng Liu$^{3}$, and Mohammad M. Ghassemi$^{1}$
\thanks{$^{1}$Department of Computer Science,
        Michigan State University, 428 S Shaw Ln, East Lansing
        {\tt\small sadiyasa@msu.edu, ghassem3@msu.edu}}%
\thanks{$^{2}$Department of Electrical \& Computer Engineering,
        New York University Abu Dhabi, Abu Dhabi
        {\tt\small tuka.alhanai@nyu.edu}}%
\thanks{$^{3}$Department of Psychology,
        Michigan State University, 316 Physics Rd, East Lansing
        {\tt\small tsliu@msu.edu}}%
\thanks{\makebox[\columnwidth]{978-1-7281-6215-7/20/\$31.00 \copyright2020 IEEE \hfill}}
}

\begin{document}

\maketitle

\thispagestyle{empty}
\pagestyle{empty}

\begin{abstract}


Electrode ``pop" artifacts originate from the spontaneous loss of connectivity between a surface and an electrode. Electroencephalography (EEG) uses a dense array of electrodes, hence \textit{popped} segments are among the most pervasive type of artifact seen during the collection of EEG data. In many cases, the continuity of EEG data is critical for downstream applications (e.g. brain machine interface) and requires that popped segments be accurately interpolated. In this paper we frame the interpolation problem as a self-learning task using a deep encoder-decoder network. We compare our approach against contemporary interpolation methods on a publicly available EEG data set. Our approach exhibited a minimum of $\sim 15\%$ improvement over contemporary approaches when tested on subjects and tasks not used during model training. We demonstrate how our model's performance can be enhanced further on novel subjects and tasks using transfer learning. All code and data associated with this study is open-source to enable ease of extension and practical use. To our knowledge, this work is the first solution to the EEG interpolation problem that uses deep learning.
\end{abstract}

\section{INTRODUCTION}

\copyrightnotice
Electroencephalography (EEG) devices have become increasingly popular in recent years and are used in a wide range of applications. Naturally, the medical applications of EEG are centered on neurological diagnosis, but EEG has proven useful for other problems in healthcare domain \cite{Selvam_2011, Xing2018}. Moreover, the use of EEG devices extends far beyond the medical domain; novel applications of EEG may be found in wide a variety of fields including advertising \cite{Vecchiato_2011}, education\cite{Siripornpanich_2018}, entertainment \cite{Bonnet_2013}, and security \cite{Khalifa_2012}.    

A fundamental challenge of EEG data is the low signal to noise ratio. Different sources contribute to this noisiness but, in general, they can be categorized as either movement artifacts or electrode artifacts. The most common, and particularly persistent, electrode artifact is the electrode \textit{``pop"} \cite{Walczak_2009, Louis_2016}. These artifacts result from abrupt changes in impedance, usually due to a loose electrode or bad conductivity. Furthermore, these artifacts are difficult to avoid because, even if the greatest care is taken when applying electrodes, the most minor subject movement or change in perspiration can cause the electrode to ``pop''. 

A common solution to EEG \textit{``pops"} is to interpolate the missing segments using recordings from nearby electrodes \cite{GabardDurnam_2018}. In practice, this interpolation is most commonly performed using \textit{eeglab}, which contains a tool for spherical interpolation \cite{Perrin_1989,Ferree_2006}. Within the last few years, alternative interpolation methods reporting improved performance have been proposed: Petrichella \textit{et al.} proposed a euclidean inverse distance method \cite{Petrichella_2016} while Courellis \textit{et al.} demonstrated an interpolation approach that (while also being based on an inverse distance calculation) used geodesic lengths and electrode localization to extract more exact channel locations, thereby performing more accurate interpolation. \cite{Courellis_2016}. Effective interpolation is a necessary preliminary step to any subsequent preprocessing or formal analysis of the EEG, including Independent Component Analysis (ICA). As noted by Ullsperger \textit{et al}: ``\textit{activity from bad channels should be removed before ICA decomposition, as it can massively deteriorate otherwise good decomposition results.}" \cite{ullsperger2010simultaneous}

One shortcoming of these previous solutions is their dependence on knowledge of the precise locations of the electrodes (i.e. electrode localization / registration), which are not collected in most practical settings. Furthermore, the existing methodologies assume that the incidence and specific characteristics of the \textit{``pops"} are similar across both subjects and tasks (i.e. ``one-size-fits-all"). Furthermore, as far as the authors are aware, none of the studies surveyed for the purposes of this work provided publicly available software repositories to enable practical use, reproduction of their methodologies, or ease of extension.

To address the aforementioned challenges, we propose a novel electrode interpolation framework using representation learning. Our method autonomously identifies the spatio-temporal properties of EEG data measured at a set of electrodes, that predict the values of a given neighbor to those electrodes. Our model, which has been made publicly available \footnote{\url{https://github.com/sari-saba-sadiya/EEG-Channel-Interpolation-Using-Deep-Encoder-decoder-Networks}}, can be used ``out of the box" to more effectively interpolate EEG for any missing channel, at any time. 

One important advantage of our model over existing approaches is its amenability to transfer learning: the ability to easily fine tune it using clean data from a novel subject or EEG experiment. This property of our model allows for interpolation that is tailor-made to the specific task and subject at hand, enabling the model to learn even idiosyncratic relations in new data.

To determine the usefulness of our method we evaluate the model on unseen tasks and subjects with and without further tuning. To summarize, our main contributions are:
\begin{itemize}
\item We propose and implement a new framework for EEG channel interpolation using encoder-decoder deep representation learning.
\item We compare our method against contemporary algorithms for channel interpolation.
\item We make our code publicly available for the benefit of the community, and demonstrate how it may be tuned to novel subjects and tasks using transfer learning.
\end{itemize}

\section{Related Work}

\subsection{Current Interpolation Solutions} \label{Current Interpolation Solutions}

The Spherical interpolation method (as implemented by \textit{eeglab}) was first proposed by Perrin \textit{et al.} in their 1989 work \cite{Perrin_1989} and remains the most widely-used interpolation solution \cite{Perrin_1989,Ferree_2006,Anthony1993,Petrichella_2016}. More recent work describes improved interpolation methods by, for instance, using ellipsoid geodesic lengths \cite{Courellis_2016} to better estimate the distances between electrodes. A weighted signal reconstruction scheme that favors electrodes closer to the location of the missing channel is then used to achieve higher accuracy. This however requires that the electrode positions be digitally registered so that an ellipsis can be fitted to the shape of the subject's head. That ellipsis is then used when calculating the distances between electrodes. Unfortunately, most data tends not to provide the specific channel locations. 

\subsection{Artifact Detection Using Neural Networks}
EEG is a signal with spatial structure that unfolds in time. Convolutional neural networks (CNNs) are an obvious candidate for the EEG interpolation problem because they naturally capture hierarchical spatio-temporal relationships. 

A few recent papers have leveraged CNNs for EEG classification \cite{Ghassemi_2019,McCartney_2019}. Previous works have also used CNNs to annotate EEG wave-forms for the presence of artifacts: Nejedly \textit{et. al.} framed artifact detection as a classification problem and used CNNs to robustly annotate eye blink segments \cite{Nejedly2018}. However, while the use of deep learning approaches for artifact \textit{detection} has shown promising results, there is relatively less work on the use of deep learning to \textit{reconstruct} artifact-ridden data segments. 


Finally, recently researchers working with EEG data started to use \textit{generative} rather than \textit{discriminative} neural networks. For instance \cite{Corley2018} used generative auto-encoders (GANs) to specially up-sample EEG data by interpolating non-existing channels. While not interchangeable, this is a similar task to channel interpolation. However, no previous work was validated on either subjects or tasks that did not appear during training. Moreover, to the best of the authors' knowledge no previous work tested how transfer learning could facilitate further fine tuning on specific data-sets, or even made the trained model available. We hope that our rigorously tested ready to use model will allow for wider access to state-of-the-art machine learning techniques.      




\section{Data}

\subsection{Data Collection}
\begin{figure*}[t]
\centering
  \includegraphics[width=1\textwidth]{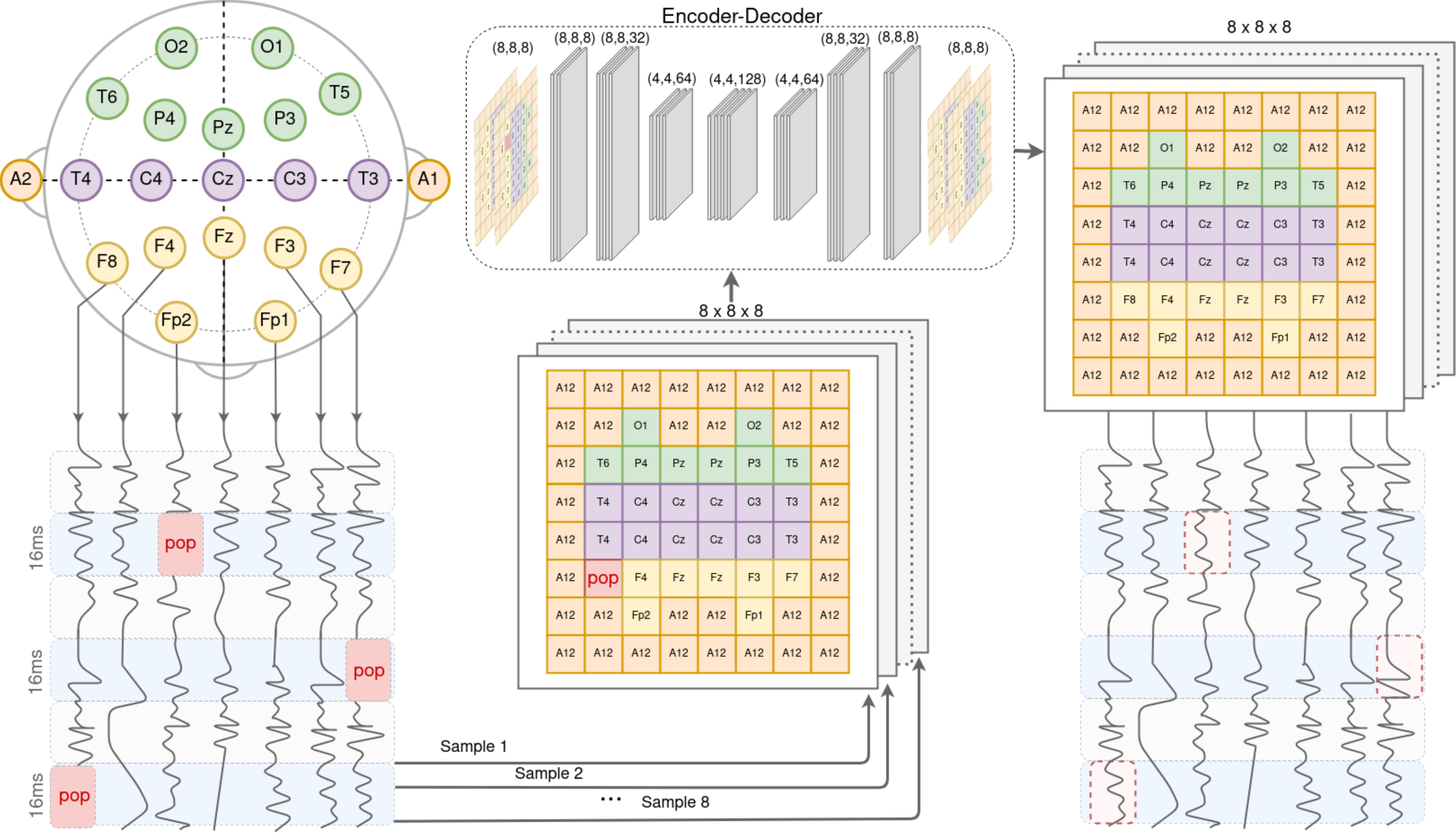}
  \caption{A diagram of our framework. The EEG data is first segmented the $500 Hz$ data into $16ms$ epochs, each segment is then mapped to a $5x5$ matrix that roughly reflects the spatial locations of the EEG electrodes (e.g. \textit{F7} is located in position 1,1 of the matrix). The electrodes at the sagittal and median planes were duplicated and the tensor was padded with the linked ear channel data to create a $8\times 8 \times 8$ tensor per epoch.  This data serves as the input to an encoder-decoder model. Finally, the output is transformed back into a signal.}
  \label{fig:digram}
\end{figure*}

The data in this study was collected from the recently published \textit{EEG During Mental Arithmetic Data Set} \cite{Zyma_2019}. The data consists of 24 subjects performing two tasks: a \textit{resting state task}, and a \textit{mental arithmetic task}. EEG data was collected as subjects performed the tasks using the 10-20 international system with the linked ears serving as a reference electrode (see Figure \ref{fig:digram}, left). The sampling rate was $500 Hz$.  Each \textit{resting state task} lasted $180$ seconds while each \textit{mental arithmetic task} lasted for $60$ seconds; hence, the total number of samples was $90,000$ and $30,000$ for each \textit{resting state} and \textit{mental arithmetic} task respectively. 

We segmented the data into $16 ms$ ($8$ samples) epochs. Hence, each subject ended up with  $11,250$ and $3,750$ epochs for the \textit{resting state} and \textit{mental arithmetic} tasks respectively.




\begin{figure}[t!]
\centering
  \includegraphics[width=0.5\textwidth]{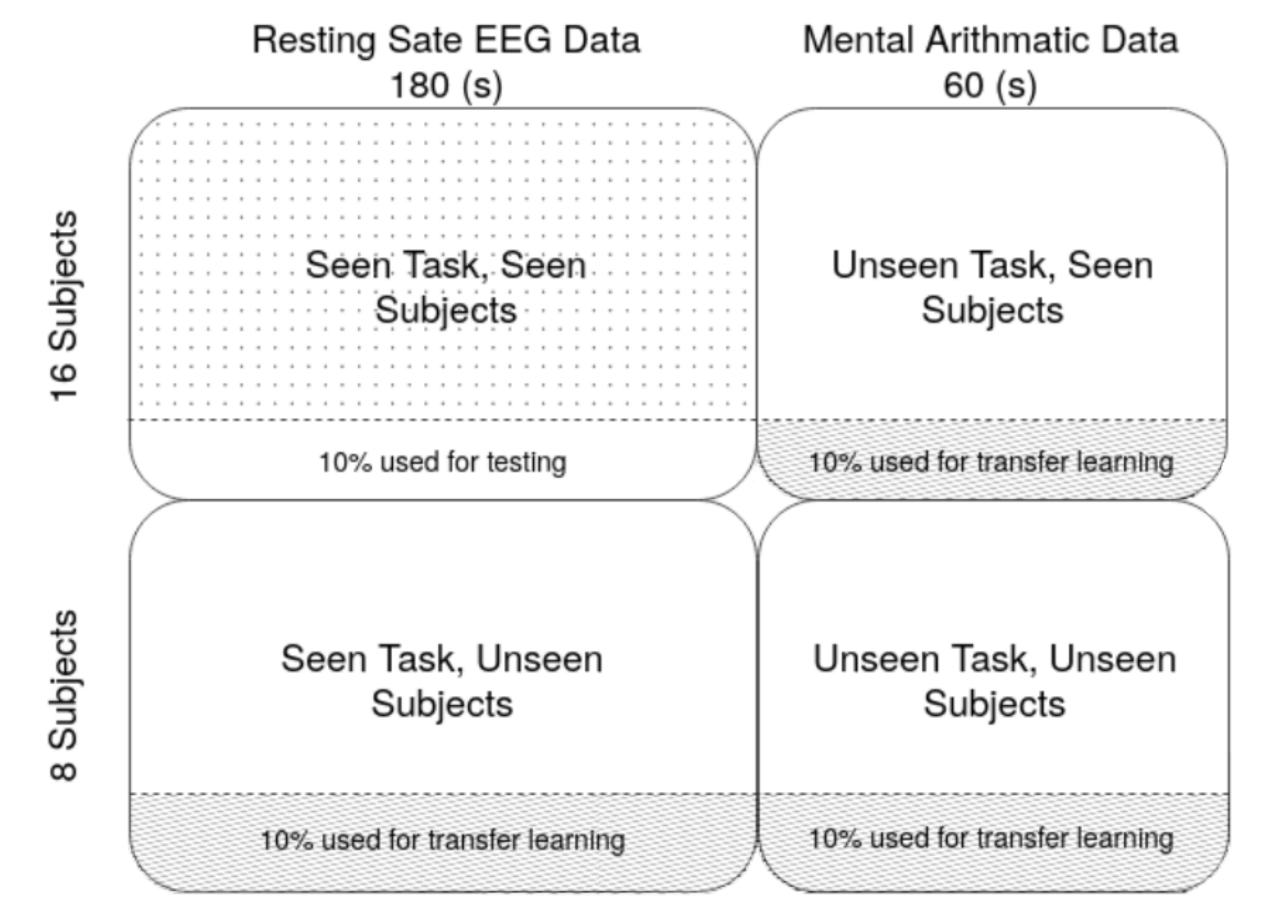}
  \caption{The four-way split of the data into partitions. The dotted area (top left) was used to train our main algorithm. The areas with the diagonal shading were used for transfer learning. The areas with white background were used to test our algorithm. Each shaded area was used separately to tune the model for testing on the remaining $90\%$ of the data in its own partition.}
  \label{fig:data_split}
\end{figure}

\subsection{Data Partitioning}
In Figure \ref{fig:data_split}, we illustrate how the data was partitioned for the training and evaluation of our method. Training data from the \textit{resting state task} was used for model development while data from the \textit{mental arithmetic task} was held out for intra-task evaluation. We further partitioned data from both tasks into training subjects (67\% of subjects) and evaluation subjects (33\%). This resulted in the following four partitions of the data, which we will refer to later when discussing our results:``\textit{Seen Task, Seen Subjects}" (n=16 subjects),``\textit{Seen Task, Unseen Subjects}" (n=8), ``\textit{Unseen Task, Seen Subjects}" (n=16), and ``\textit{Unseen Task, Unseen Subjects}" (n=8). 

The ``unseen" data sets contain some deviation from the data the model was trained on, and are thus a good test for the generalizability of the method. Because we are utilizing transfer learning $10\%$ of the data for all unseen sets was held-out for tuning, to test the impact of this additional context on the model's performance.

In general, the fundamental difference between the two tasks is crucial to our evaluation. Researchers will often not have enough data to train neural networks "from scratch", not to mention training a neural network often requires exploration of a hyper-parameter space that may consume significant temporal (and financial) resources. This is especially true for deep learning frameworks that have strong tendencies to over fit and produce remarkable results for a specific data set while failing to generalize to other contexts. With this in mind, it is important that our models achieve good results on both unseen tasks and subjects to enable their continued development and utility within the greater research community. We therefore structured our data to assess its ability to generalize across tasks and subjects.



\section{Methods}

\subsection{Pre-processing} \label{Pre-processing}
To begin, all EEG data were Z-scored at the subject-level (i.e. converted to a zero mean, and unit variance representations). Deep networks require a large volume of training data; hence, we supplemented our training data by transforming each subject's EEG data into 10 distinct \textit{pseudosubjects}. Each pseudosubject's data was an elementwise addition of the Z-scored subject's EEG data and random draws from a Gaussian distribution with ($\mu=0$, $\sigma=0.05$). The pseudosubject's (already normalized) EEG data was then Z-scored again following the introduction of the noise. The utility and validity of this data augmentation approach for EEG research has been demonstrated in prior work \cite{ghassemi2018life}. Next, a simple transformation was applied to project each sample of EEG data from a spherical channel representation onto a quantized two dimensional surface represented by a $5\times 5$ matrix (Figure \ref{fig:digram}, panel 2). 

Finally, EEG data was epoched into $16ms$ segments, with no overlap across segments \footnote{the average duration of a pop artifact exceeds 1 second, hence 16 ms is more than sufficient for reconstruction.}. This resulted in a $5$ channels $\times 5$ channels $\times 8 $ samples tensor. The electrodes at the sagittal and median planes (the central electrodes) were then duplicated and the tensor was padded with the linked ear channel data to create a $8\times 8 \times 8 $ tensor. This manipulation of input size is common place in deep learning and is mainly the result of networks being optimized to work with input sizes that are powers of two \cite{Lin_2015}. This $8\times 8 \times 8 $ tensor formed a single sample of input data, from the perspective of the network when training. 

To create training data, we iterative occluded each of the $19$ non reference electrodes (All electrodes except $A1$ or $A2$, see Figure \ref{fig:digram}). Thus each $8\times 8 \times 8 $ tensor became the prediction target for $19$ input tensors, each with one distinct occluded channel.



\subsection{Proposed Approach}
\subsubsection{An Encoder-Decoder model for EEG Interpolation}
Inspired by research on image inpainting we deployed an encoder-decoder model for EEG interpolation. Image inpainting is a classical problem in computer vision: given a corrupted image the aim is to complete or ``fill in" missing pixels. This is a similar problem to electrode interpolation.  

Encoder-decoder models are the combinations of two networks that are trained simultaneously: the encoder first learns a lower dimensional embedding of the data, and the decoder attempts to recover the original data from the embedding. Encoder-decoder networks are a popular tool in image inpainting  \cite{pathak_2016,yang2019lafin,wang2019srn}, so much so that this technique is now leveraged for image compression as selective removal of pixels might greatly enhance compression ratios \cite{Baig_2017}. 

\begin{figure}[ht]
\centering
  \includegraphics[width=0.49\textwidth]{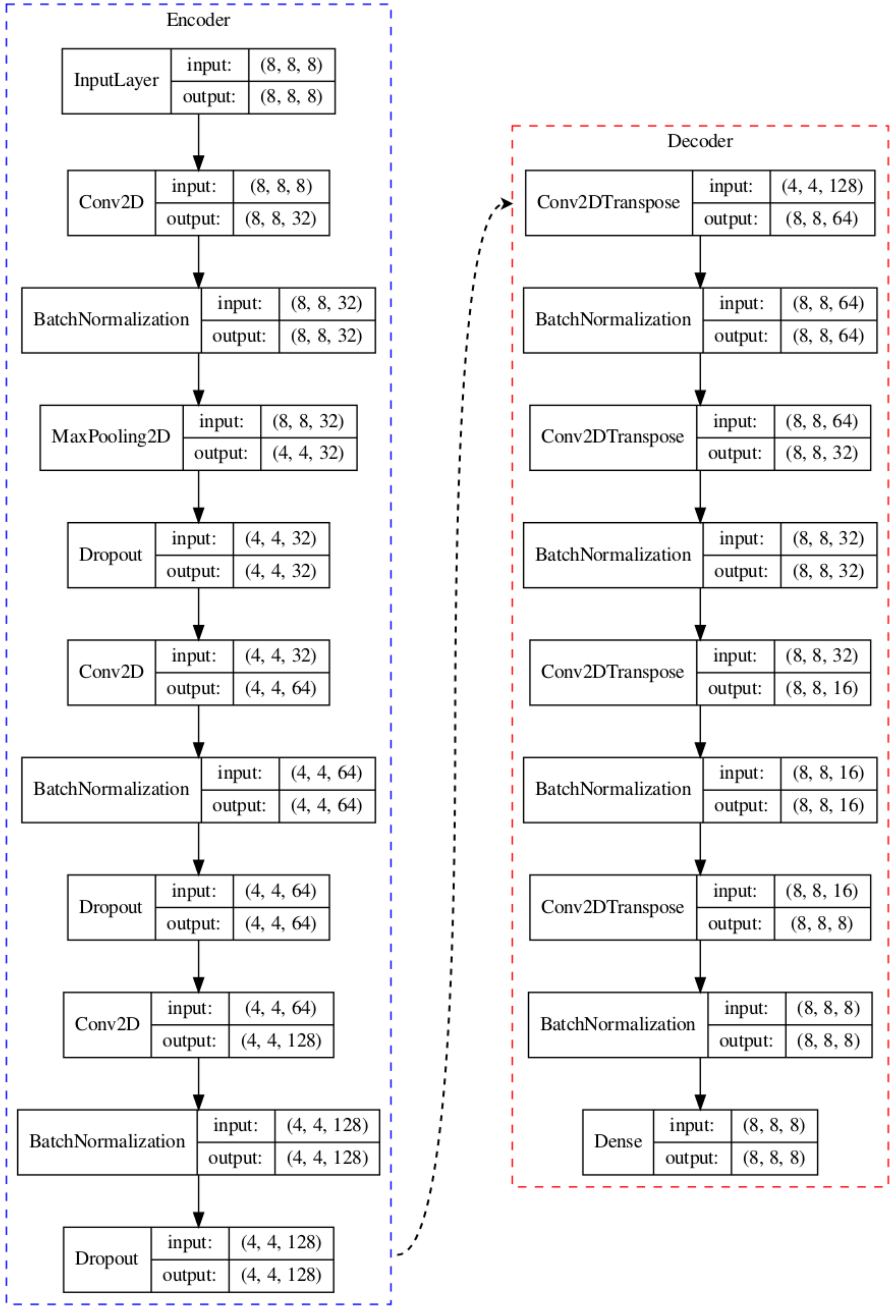}
  \caption{Our network, the dashed black arrow denotes the $4\times4\times 128$ embedded data tensor. This embedded representation is passed form to the decoder which reconstructs the original input sans the occlusion.}
  \label{fig:encoder_decoder}
\end{figure}

We determined the optimal topological configuration of our encoder-decoder network via a random search of the network hyper-parameter space \cite{bergstra2012random}. The tested topologies varied in the number of convolution layers, the existence of max-pooling, dropout, and batch normalization layers after each convolution layer, and whether the decoder was based on transposed convolution layers or simple up-sampling with convolution. More specifically, we trained 300 distinct architectural configurations of the encoder-decoder networks, and retained the configuration that best generalized within a held out subset of the training data itself. The best network was then used after training for our transfer learning evaluation. The code to run this search in the topological space, as well as the trained winning algorithm before and after the transfer learning tuning is available online. The optimal topology is shown in Figure \ref{fig:encoder_decoder}, and discussed in Subsection \ref{Model Hyper-parameter Optimization}.



\subsubsection{Subject+Task Enhancement via Transfer Learning}
Transfer learning experiments were carried out by taking the model trained on the original data set and tuning it on a small subset ($10\%$) of the testing data. Realistically, it is highly likely that a small sample of clean data will be available for a researcher to use when tuning our model. 

To assess performance enhancements associated with transfer learning, we held out $10\%$ of each data partition (See Figure \ref{fig:data_split}) and tuned our network for $100$ epochs. These numbers were intentionally small as to showcase how even minimal training that can be easily completed on non specialized hardware and using very little data can lead to significant improvements. By tuning the model for specific subjects and tasks, we assessed the flexibility and practical extensibility of our proposed approach.


\subsection{Methodological Baselines}  \label{subsection:baselines}


For our baseline we implemented the three methods described in the Related Work section \cite{Perrin_1989,Petrichella_2016,Courellis_2016}. 

\subsubsection{The Euclidean Baseline} 

Both \cite{Petrichella_2016,Courellis_2016} suggest methods that employ an inverse distance metric where the interpolated channel $\hat{s_i}$ is calculated using the following equation:

$$ \hat{s_i} = \frac{\sum_{j\neq i} w_{ij} s_j }{\sum_{j\neq i} w_{ij}} , w_{ij} = \frac{1}{d_{ij}^p} $$

where $p$ is the power parameter. The variable $d_{ij}$ represents the distance between electrodes $i$ and $j$. The original channel $j$ is represented by $s_j$. The power parameter is an integer (usually between 2 and 5) that is set using a small amount of data; while it is usually set to be the same value across a given data-set where interpolation is happening, we optimized the power parameter separately for each baseline and data-set to maximize the performance of the baselines.


The calculation of the distance,  $d_{ij}$,  is the main difference between the two baselines. The fist euclidean baseline (\textit{EUD}) uses a simple euclidean distance formula. This distance calculation is done using the generic electrode positions in space that are always available for every cap. 

\subsubsection{The Geodesic Baseline}
The Geodesic Length baseline (\textit{EGL}) is also based on the inverse distance equation. However, instead of using euclidean distances for $d_{ij}$ the geodesic length is calculated. The geodesic distance is calculated using the \textit{Vincenty} algorithm which was originally used in geodesy to calculate the distance between points on the surface of a spheroid. The method is iterative and not theoretically guaranteed to converge. Previous work have demonstrated that interpolation calculated using this method outperforms simple euclidean interpolation \cite{Courellis_2016}. However, as previously discussed, this was tested by the original baseline works when specific electrode locations were available \cite{Courellis_2016}. It should therefore be expected that results obtained for the \textit{EGL} may be lower than those reported in previous studies.

\subsubsection{The Spherical Splines Baseline}
Finally, we also followed the \textit{eeglab} MATLAB implementation of spherical splines method (\textit{SS}) \cite{Perrin_1989}. According to this implementation at each point in time the value of of the interpolated channel $\hat{s_i}$ can be approximated using the equation:

$$ \hat{s_i} = c_0 +  \sum_{j\neq i}  c_j g\left( cos(\theta_{i,j}) \right) $$

Where $\theta_{i,j}$ is the angle between the electrode locations $i$ and $j$. Instead of calculating the angle, given the positions of the electrodes in space $p_i=(x_i,y_i,z_i)$ it is possible to directly calculate the cosine value: $ \cos(\theta_{i,j})=\frac{x_i \cdot x_j}{ \norm{x_i} \cdot \norm{x_j} }$. The function $g(x)$ is defined as the sum of the series:

$$ g(x) = \frac{1}{4\pi} \sum_{n=1}^{\infty} \frac{2n+1}{n^m (n+1)^m} P_n(x)$$

Where $P_n$ is the Legendre polynomial and following \cite{Perrin_1989} we set $m=4$. Additionally, following the \textit{eeglab} implementation we limited the infinite sum to the first seven values. Finally, the coefficients $C=(c_1,c_2,\dots,c_n)$ are set to be the solution for the system of equation $GC+T c_0 = S$ with the constraint $T'C=0$ where $G_{k,l}=g\left( cos(\theta_{k,l}) \right)$, $S=(s_1,s_2,..,s_n)$ and $T$ is a vector of ones. The interpolated channel is excluded from the calculations of $G$ and $S$.


\subsection{Model Evaluation Approach}
The selected baseline approaches \cite{Courellis_2016} used the averaged normalized mean square error (ANMSE) as the main evaluation measure. The normalization of the mean square error is used to prevent a specific channel's performance from skewing the results, in case of a bad reconstruction. Having Z-scored the data however all channels are guaranteed to have the same mean amplitude. Therefore we do not normalize our mean square error results. Hence our final evaluation is:


$$AMSE = \frac{1}{M} \sum_{j=1}^{M} \left( \frac{\sum_{i=1}^{N} (s_{i} - \hat{s}_i)^2}{N} \right)_j $$

Where $N$ is the number of channels (19 in our specific case). $M$ is the number of samples. Note that the expression for mean reconstruction error (the inner average) changes for every sample. $s_{i}$ and $\hat{s}_i$ are as previously defined. This measure was used both for optimizing the power parameter for the different baselines (see Section \ref{subsection:baselines}) and calculating the final results presented momentarily.

\begin{table*}[t!]
\begin{center}
\caption{Comparison between Encoder-decoder model and baselines using Averaged mean square error (AMSE); \textit{lower is better}. Note that for transfer learning (last row) the training data set differed on each column. There was no transfer learning for the  \textit{Seen Task, Seen Subject} partition as this is the original data used to train the model. SS: spherical splines baseline EGL: geodesic length calculation; EUD: euclidean baseline. The best result is bolded and percentage of improvement over the most competitive baseline is given.}
\footnotesize
\centering
\label{tab:main-results}
\begin{tabular}{|M{3.6cm}||M{2.8cm}|M{2.8cm}|M{2.8cm}|M{2.8cm}|}
\hline
\specialcell{Interpolation\\ Method} & \specialcell{Seen Task,\\ Seen Subjects \\ (\% Improvement)} & \specialcell{Seen Task,\\ Unseen Subjects} & \specialcell{Unseen Tasks,\\ Seen Subjects} & \specialcell{Unseen Task,\\ Unseen Subjects} \\
\hline
\hline
\specialcell{SS Baseline} & $0.728$ & $0.694$ & $0.8238$ & $0.779$\\
\hline
\specialcell{EUD Baseline} & $0.5215$ & $0.561$ & $0.665$ & $0.566$\\
\hline
\specialcell{EGL Baseline} & $0.585$ & $0.501$ & $0.566$ & $0.622$\\
\hline
\specialcell{Our Encoder-Decoder Model} & {\boldmath $0.446$} (14.47\%) & $0.478$ &  $0.552$ & $0.465$ \\
\hline
\specialcell{Our Encoder-Decoder Model +\\ Transfer Learning} & --- & {\boldmath  $0.392$} (21.75\%) & {\boldmath  $0.439$} (21\%)& {\boldmath $0.446$} (19.78\%)\\
\hline
\end{tabular}
\end{center}
\end{table*}

\section{Results}



\subsection{Model Hyper-parameter Optimization} \label{Model Hyper-parameter Optimization}

After exploring the topological space by testing different network architectures (using the \textit{Seen task, Seen subjects} data, see Figure \ref{fig:data_split}), the best performing network is visualized in Figure \ref{fig:encoder_decoder}. This network consisted of a simple encoder with three convolution layers and one max pooling layer, as well as four transposed convolutions in the decoder. Additionally there was a dropout and a batch normalization layer after each convolution in the encoder. All results described in this section were achieved using this particular architecture.


\begin{figure}[t!]
\centering
  \includegraphics[width=0.45\textwidth]{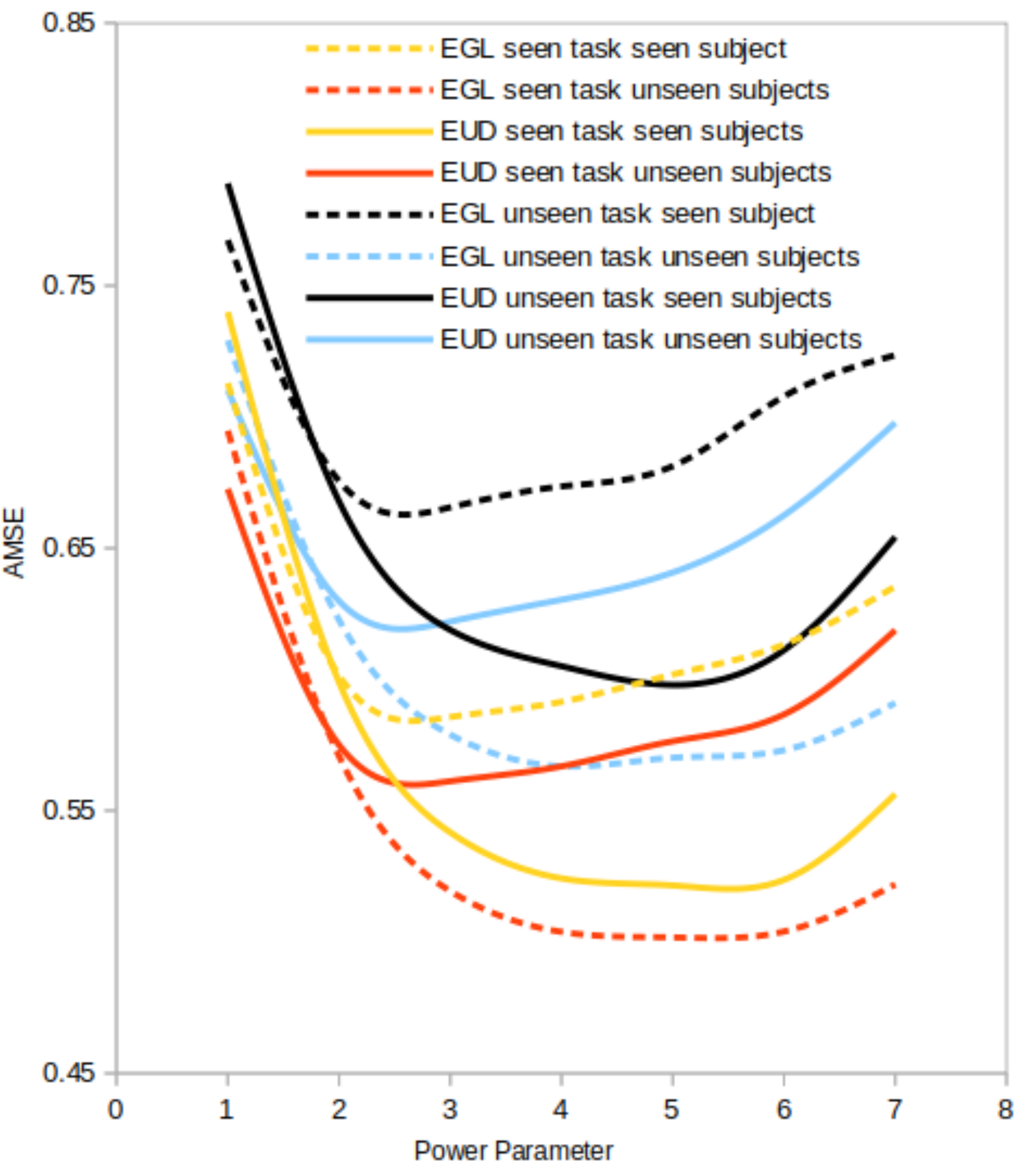}
  \caption{Power parameter optimization to maximize the performance of the baseline approaches. For each of the four data partitions and for the two, EUD and EGL (solid and dashed lines respectively), methods. EGL: Geodesic Length calculation; EUD: euclidean baseline; AMSE: Averaged mean square error}
  \label{baseline_opt}
\end{figure}

\subsection{Baseline Power-parameter Optimization}
For our evaluation to be extra rigorous. we optimized the power parameter for each baseline data-partition configuration separately. In Figure \ref{baseline_opt}, we illustrate the results of our power parameter optimization for the baseline methods. As seen in the Figure, the optimal power parameters were comparable with those reported in previous literature (between 2 and 5) \cite{Courellis_2016}. All the results that are reported in this section were for the optimized baseline on the specific data set being discussed. The spherical splines baseline has no analogous parameter we can optimize.

\begin{figure*}[t!]
\centering
  \includegraphics[width=1\textwidth ]{ecr_figure3_3.pdf}
  \caption{An exemplary 48ms reconstruction of the EEG data for Subject $0$ \textit{resting state task} for channel \textit{P4}. The original channel data was removed and interpolated using best performing baseline (geodesic length calculation, in blue) and our method (in red).}
  \label{fig:reconstruction}
\end{figure*}


\subsection{Main Result}

In Table \ref{tab:main-results}, we compare the results of our proposed approach against the baselines for the EEG interpolation task on the test sets. The baseline methods are highly unstable, giving a high variability in performance relative to our approach.



The Encoder-decoder model consistently outperformed the baselines by at least $10\%$. Moreover, by utilizing transfer learning the network was able to improve it's accuracy even with minimal additional data and training time. 

Interestingly, in contrast to results reported in the literature, the \textit{EGL} method did not clearly outperform the \textit{EUD} baseline \cite{Courellis_2016}. This might be due to our data not having the precise electrode locations in contrast to previous research (see Subsection \ref{On the challenges of electrode localization} in the discussion). 

Another interesting result is the pattern of improvements after transfer learning. As can be seen in the last row of Table \ref{tab:main-results}, the biggest improvement was for \textit{Unseen task, Seen subjects} data. This hint that there was more variability between EEG data from different tasks compared to data from different subjects. Additional testing will be needed to verify this hypothesis. However, this can be seen as a compelling argument in favor of flexible models that can be tuned for the specific data the researcher is working with.  

Finally, we also extracted the delta ($0.5$-$4 Hz$), theta ($4$-$8 Hz$), alpha ($8$-$12 Hz$), beta ($12$-$30 Hz$), and gamma ($30$-$100 Hz$) bands and tested the models performance for each band separately. Our method significantly improved over the baselines method in all bands. This is crucial as different bands have different functions (for instance, the theta band is especially responsive during observation and memorization tasks \cite{Vecchiato_2011}). Hence for an interpolation method to be useful the reconstruction fidelity must be consistent across all frequency bands. In the interest of brevity we will not present the results for all these bands separately. The code to extract the sub-bands is also available online.     



\subsection{Performance on Exemplary Data}
In Figure \ref{fig:reconstruction}, we present an example of our method's interpolation on an exemplary portion of the data, compared against the baselines. As shown in the figure, the best baseline reconstruction contains voltage fluctuations that do not appear in the original signal, or the one reconstructed using our method. These fluctuations were quite common in baseline reconstructions. We speculate that our method might have learned to not only the optimal weights to use to approximate the occluded channel, but also a more complicated relationship that enables our method to suppress potential artifacts that are localized to one electrode and therefore do not effect the original electrode that is being reconstructed. All things being equal, this is evidence that our framework was able to learn the nuanced relationships between electrode measurements that are not captured by baseline approaches.

\section{Discussion}







Our work used a deep encoder-decoder model to tackle the problem of EEG channel interpolation. While discriminative frameworks are able to only detect and label bad data segments, our results demonstrate that a generative approach can reconstruct the missing channel with high fidelity to the original signal. The success of our method suggests that deep learning can capture complex relationships between electrodes that are not sufficiently expressed by the relatively simple inverse distance calculations predominant in contemporary solutions.

\subsection{On Self-supervised Learning}
Data labeling is often a tenuous and resource consuming process. Unfortunately, training deep learning models often requires extensive data collection and labeling efforts. Therefore, deep learning researchers have recently began to focus on finding ways to mitigate the need for labeled data. As we showed in this study, one approach to mitigate this is to frame problems as self-supervised learning tasks\footnote{For a particularly interesting primer on self-supervised learning see \href{https://drive.google.com/file/d/1r-mDL4IX_hzZLDBKp8_e8VZqD7fOzBkF/view}{Yann LeCun's Keynote lecture at the Thirty-Forth annual meeting of AAAI.}}. Specifically, our work is a special case of a popular self-supervised learning task: the prediction of occluded parts of data from visible ones. By using this framing we were able to circumvent a common hurdle faced by deep learning approaches. 


\subsection{On the Challenges of Electrode Localization} \label{On the challenges of electrode localization}
As discussed previously, prior research that compared different interpolation methods used electrode localization to extract exact channel locations for each specific subject. While generic and imprecise locations are always available, electrode localization methods attempt to alleviate the noisiness inherent to EEG by providing exact electrode locations. This localization can be done in many ways; one expensive option is to equip EEG caps with spatial sensors, or motion capture sensors \cite{Courellis_2016,Reis2015}. Other methods that require less specialized hardware including a simple DSLR camera \cite{Clausner2017} and Kinect with an Neural Network \cite{Gessert2019TowardsDL}. However, despite these recent advances, electrode localization remains uncommon. For instance, no EEG data set in physionet\footnote{\url{https://physionet.org/about/database/\#ecg}} or gigadb\footnote{\url{http://gigadb.org/search/new?keyword=eeg}} contain an EEG database with electrode localization. Therefore, and to ensure our method is applicable to the vast majority of databases, the data we used also did not include electrode localization \cite{Zyma_2019}. A possible future work could incorporate location data into the deep learning framework.


\subsection{On Baseline Approaches}
It is worth noting that there are multiple other interpolation methods such as the nearest neighbors method, planar-spline technique \cite{Anthony1993}. We selected the baselines methods described in Subsection \ref{Current Interpolation Solutions} as they were the most contemporary approaches on the topic. Furthermore, the performance improvement of our model are especially impressive considering that the \textit{EUD} and \textit{EGL} baselines were optimized to maximize their performance on each and every separate partition of the data.

The \textit{SS} method requires a system of equations to be solved for each and every time point. This is not a trivial requirement as it necessities complex calculations. This demand renders the \textit{SS} method ill-suited for any online interpolation, and by extension many BCI applications \cite{Bonnet_2013,Lin_2013,Xing2018}. In contrast to the taxing nature of the training procedure, piping data foreword in neural network is computationally cheap. Therefore our approach could potentially satisfy a growing need for accurate interpolation from online data.     

\subsection{On Transfer Learning}
Transfer learning involves training a model on a problem similar to the one being solved. This is especially useful when only scarce data is available for the problem being solved, hindering the training of the model. While transfer learning is possible for many machine learning algorithms such as Bayesian networks and Markov chains, this technique became essential to deep learning especially due to its reliance on huge amounts of training data. Transfer learning is considered to be essential for the success and ubiquity of neural networks \cite{Ng2016}. Our work for instance would be considerably less useful if it required every researcher to train the neural network from scratch, or if the results on data-sets that the model was not trained on were considerably worse.

\section{Conclusion and Future Work}


With the increasing prevalence of EEG devices, there is a need for methodologies that better address common EEG artifacts. In this work, we developed a deep encoder-decoder based method to interpolate EEG segments impacted by the most common EEG artifact: the electrode \textit{``pop"}. We demonstrated that our method improved EEG reconstruction performance compared to existing approaches, and that our method generalized well to unseen tasks and subjects. 


Future work will extend this method to tackle other kinds of electrode artifacts. Moreover, an end-to-end system that automatically detects artifacts and replaces the corrupted data with an interpolated reconstruction of the original might be of particular interest to the community. 


\section*{ACKNOWLEDGMENT}
We would like to thank the authors of \textit{EEG During Mental Arithmetic Data Set} \cite{Zyma_2019} for making their data available.\\
This work was partially supported by the Michigan State University Graduate Office Fellowship.

\addtolength{\textheight}{-10cm}   





\bibliographystyle{ieeetr}
\bibliography{bib}

\end{document}